\documentclass[12pt,letterpaper,english]{article}
\usepackage{amssymb,amsmath}
\include{epsf}
\usepackage{slashed}

\usepackage{graphicx}
\usepackage{amsfonts}

\newcommand{\e}{{\rm e}}
\newcommand{\ii}{{\rm i}}

\newcommand{\eqn}[1]{(\ref{#1})}

\def\appendix#1{\addtocounter{section}{1}\setcounter{equation}{0}
\renewcommand{\thesection}{\Alph{section}}
\section*{
\thesection\protect\indent \parbox[t]{11.715cm} {#1}}
\addcontentsline{toc}{section}{Appendix\thesection\ \ \ #1} }
\newcommand{\complex}{{\mathbb C}} 
\newcommand{\real}{{\mathbb R}} 

\newcommand{\tr}[1]{\:{\rm tr}\,#1}
\newcommand{\Tr}[1]{\:{\rm Tr}\,#1}
\newcommand{\be}{\begin{equation}}
\newcommand{\ee}{\end{equation}}
\newcommand{\beq}{\begin{equation}}
\newcommand{\eeq}{\end{equation}}
\newcommand{\bea}{\begin{eqnarray}}
\newcommand{\eea}{\end{eqnarray}}
\def\beqa{\begin{eqnarray}}
\def\eeqa{\end{eqnarray}}
\def\nn{\nonumber}
\newcommand{\del}{\partial}

\newcommand{\eq}{\begin{equation}}
\newcommand{\eqa}{\begin{eqnarray}}
\newcommand{\en}{\end{equation}}
\newcommand{\ena}{\end{eqnarray}}

\newcommand{\X}{\mathfrak X}

\def\R{{\mathbb R}} \def\C{{\mathbb C}} 
 \def\one{\mbox{1 \kern-.59em {\rm l}}}

\def\a{\alpha} 
  \def\la{\lambda}

\def\cA{{\cal A}}      
\def\cH{{\cal H}}    \def\cL{{\cal L}}  
  \def\cQ{{\cal Q}}

\def\obar{\overline}

\begin{document}
\begin{titlepage}
\begin{flushright}

\baselineskip=12pt
DSF/3/2010\\
ICCUB-10-006\\
UWTHPh-2010-1\\ \hfill{ }
\end{flushright}

\begin{center}

{\Large\bf Noncommutative gauge theory and  \\[1ex] symmetry breaking in matrix models}

\baselineskip=14pt

\vspace{1cm}

{\bf Harald Grosse$^{a}$,  {\bf Fedele Lizzi\,}$^{b,c}$ and {\bf
Harold Steinacker}$^{a}$}
\\[5mm] $^a$ {\it Department of Physics, University of Vienna\\Boltzmanngasse5, A-190 Vienna, Austria}
\\{\small \tt
harald.grosse, harold.steinacker @univie.ac.at}\\[6mm]
$^b$ {\it Dipartimento di Scienze Fisiche, Universit\`{a} di
Napoli {\sl Federico II}} and {\it INFN, Sezione di Napoli,
Via Cintia, 80126 Napoli, Italy}\\
{\small \tt fedele.lizzi@na.infn.it}
\\[5mm]
$^c$ {\it High Energy Physics Group, Dept.\ Estructura i
Constituents de la Mat\`eria and Institut de Ci\`encies del Cosmos
\\Universitat de Barcelona Barcelona, Catalonia, Spain}

\end{center}


\begin{abstract}

We show how the fields and particles of the standard model can be
naturally realized in noncommutative gauge theory. Starting with a
Yang-Mills matrix model in more than 4 dimensions, a $SU(n)$ gauge
theory on a Moyal-Weyl space arises with all matter and fields in
the adjoint of the gauge group. We show how this gauge symmetry can
be broken spontaneously down to $SU(3)_c \times SU(2)_L \times
U(1)_Q$ (resp.\ $SU(3)_c \times U(1)_Q$), which couples
appropriately to all fields in the standard model. An additional
$U(1)_B$ gauge group arises which is anomalous at low energies,
while the trace-$U(1)$ sector is understood in terms of emergent
gravity. A number of additional fields arise which we assume to be
massive, in a pattern that is reminiscent of supersymmetry. The
symmetry breaking might arise via spontaneously generated fuzzy
spheres, in which case the mechanism is similar to brane
constructions in string theory.

\end{abstract}

\end{titlepage}

\section{Introduction}

While no one knows how to describe physics at the Planck scale, there
are suggestions that it may be described by some generalization of
ordinary spaces which goes under the generic name of
\emph{noncommutative geometry}~\cite{connes,landi,ticos,Madore}.
Regardless of the details of such a construction,
the noncommutative generalization of the coordinate functions
will be some matrices which satisfy commutation relations of the type
\be
[x^\mu,x^\nu]=\ii\theta^{\mu\nu}
\ee
where $\theta^{\mu\nu}$ is a quantity of the order of the square
of Planck's length. An action is then naturally defined
as some kind of matrix model 
in terms of these noncommutative coordinates,
such as the models introduced 
in~\cite{EguchiKawai,IKKT,Alekseev:2000fd}.
These matrix models are known to
describe noncommutative gauge theory
\cite{Aoki:1999vr,Douglas:2001ba}, and contain
gravity as an emergent phenomenon \cite{Steinackeroriginal}
a la Sakharov~\cite{Sakharov, Visser}. Thus they are  promising
candidates for a quantum theory of fundamental interactions. However,
the noncommutative gauge theories obtained in this manner are
quite restrictive \cite{nogo}: only $U(n)$ gauge groups
(or possibly products thereof) are consistent,
fermions can be introduced only in the adjoint or possibly
(anti-)fundamental representation, and
the trace-$U(1)$ sector is afflicted with the notorious
UV/IR mixing 
\cite{Minwalla:1999px,Matusis:2000jf,Khoze:2004zc,Jaeckel:2005wt}.
Hence these models are often thought to be incompatible with
particle physics. 
There are proposals how 
to formulate the standard model on noncommutative 
space-time
based on different approaches such as a Seiberg-Witten expansion or
open Wilson lines  \cite{Calmet:2001na,chaichian}, which however lead to
serious drawbacks in particular for the quantizations
\cite{Wulkenhaar:2001sq,unitarity}.

The main point of this paper is to demonstrate that the 
simple matrix models for noncommutative gauge theory
may nevertheless lead to low-energy gauge theories which are
extensions of the standard model.
In particular, we show how all fermions in the standard model
with their appropriate charges can be accommodated.
The principal idea is to consider a matrix model which
describes not only the usual Moyal plane $\R^4_\theta$,
but also extra dimension encoded by additional matrices. These
matrices corresponding to extra dimensions can be equivalently
interpreted as scalar fields on $\R^4_\theta$, and can acquire
nontrivial vacuum expectation values leading
via the usual Higgs effect to spontaneous symmetry
breaking. The extra-dimensional matrices
are assumed to have a finite spectrum and no massless
modes, similar in a sense
to Connes approach to the
standard model in noncommutative geometry~\cite{connes,
ConnesLott, AC2M2}. The mechanism is essentially the same as
the generation of fuzzy extra dimensions in ordinary gauge theory
\cite{AGSZ}, cf. \cite{Aschieri:2003vy}.
The trace-$U(1)$ components and its UV/IR mixing
were understood in \cite{Steinackeroriginal}
to be part of gravity sector and are not part of the
low-energy gauge theory. This allows us to 
resolve the problems with the 
$U(1)$ sector found in previous formulations of the 
standard model on the Moyal-Weyl plane
\cite{Khoze:2004zc,Jaeckel:2005wt} based on 
(products of) $U(N)$ gauge groups.

The models we will describe below have some key features of the
standard model, mainly regarding symmetry breaking, but are not yet
phenomenologically viable, in the sense that there are still several
features which are unrealistic. However, the basic mechanism based
on spontaneous symmetry breaking of the underlying noncommutative
$SU(N)$ (resp.\ $U(N)$) gauge theory is rather general, and it is
quite conceivable that more sophisticated versions might be
realistic. In particular, we will see that a promising line of
development is to consider the internal space as fuzzy
spheres, similar as in~\cite{AGSZ}. Then the pattern which emerges
is quite similar to string-theoretical constructions of (extensions
of the) standard model
\cite{Antoniadis:2002qm,Blumenhagen:2006ci,Kiritsis}, based on
strings stretching between branes. These modes are recovered here as
bi-modules of $SU(n_i)$ subgroups of the spontaneously broken $SU(N)$
gauge group. One of the main open problems is the origin of
chirality, and we only discuss some possible avenues here. This
problem is similar as in the commutative case
\cite{Steinacker:2007ay}, and can probably be solved by invoking
more sophisticated geometrical structures such as orbifolds
\cite{orbifolds}.

This paper is structured as follows. After recalling the basic
constructions of matrix models and noncommutative gauge theory,
we discuss in section 3 the symmetry breaking of $SU(n)$
to products of $SU(n_i)$ via extra dimensions. We consider both
a simplified effective treatment involving only
the low-energy degrees of freedom, as well as a more
sophisticated realization in terms of fuzzy spheres in extra dimensions.
Section 4 contains the main results of the paper, namely
the embedding of the standard model particles and fields
in the basic matrices which are in the adjoint of
$SU(N)$, focusing on $N=7$. The electroweak symmetry breaking
is discussed in section 5, as well as the structure of the
Yukawa couplings. Here we only exhibit some qualitative aspects
and discuss possible avenues for further studies.
Parts of the present paper have been presented in the 
proceedings \cite{wroclaw}.

\section{The Matrix Model}
\setcounter{equation}{0}

We start with a Yang-Mills matrix model which involves\footnote{ The
case $D=10$ is of particular interest. In this case it is possible
to impose a Majorana-Weyl condition on $\Psi$, and the model admits
an extended supersymmetry~\cite{IKKT}. On a 4-dimensional Moyal-Weyl
background as discussed below, the model then reduces to the $N=4$
SYM on $\R^4_\theta$, which is expected to be well-behaved upon
quantization.} $D = 4+n$ matrices $X^a$ and a set of fermions:
\be
S_{YM} = - (2\pi)^{2}\frac{\Lambda_{NC}^4}{g^2}\, \Tr
\left([X^a,X^b] [X^{a'},X^{b'}] \eta_{aa'}\eta_{bb'} \,\, + \,\,
\obar\Psi  \Gamma_a [X^a,\Psi] \right) \label{YM-action-1}
\ee
where $X^a$ are infinite-dimensional hermitian matrices
$Mat(\infty,\C)$, or operators in a Hilbert space $\cH$.
$\Gamma_a$ generates the $SO(1,D-1)$  Clifford
algebra, the metric $\eta_{aa'}$
is the flat Minkowski (or Euclidean)
metric with the mostly minus choice of signs, and
$\Psi$ is a corresponding (Grassmann-valued) spinor
 taking values in $Mat(\infty,\C)$.
We introduced a scale parameter $\Lambda_{NC}$ which
will be identified with the
scale of noncommutativity below, and
$g$ will be identified as a gauge coupling constant.
This model is invariant under the symmetry
\be
X^a \to U X^a U^{-1} , \qquad U \in U(\cH) .
\ee
The equations of motion of the bosonic part of model are
\be
[X^a,[X^b,X^{a'}]]\eta_{aa'} = 0
\ee
we will discuss the fermions later. There are several solutions
for this classical equation of motion, which we will often call
\emph{vacua} in the following. Apart from the trivial one
($X^a=0$) and the case in which all of the $X$'s commute, a
relevant vacuum for our model is the ``scalar Moyal-Weyl'' vacuum:
\be
[X_0^a,X_0^b]=i\theta^{ab} \label{u1vacuum}
\ee
with $\theta^{ab}$ constant. The functions of the $X_0$'s in this
case generate an algebra isomorphic (under appropriate regularity
conditions) to the algebra of  functions on a $D$ dimensional
space multiplied with the Gr\"onewold-Moyal product. That is,
given two functions $f$ and $g$, then consider $f(x), g(x)$ as
ordinary functions on the plane, and $f(X), g(X)$ operators, then
\be
f(X)g(X)=(f\star g)(X)
\ee
with
\be
\left.(f\star
g)(x)=\e^{-\frac\ii2\theta^{ab}\del_{x^a}\del_{y^b}}f(x)g(y)\right|_{x=y}
\ee
We interpret this as the fact that the vacuum~\eqn{u1vacuum}
describes a noncommutative space where the coordinates have a
nontrivial constant commutator, the noncommutative space
$\real^D_\theta$. The bosonic part action has a gauge invariance for
the unitary elements of the algebra, since we are considering
functions of $X$ we consider these unitary elements as unitary
matrix functions $U(X)$ to which corresponds a function of $x$ which
is as usual a phase. We call this association of functions of the
matrices with functions on an ordinary space the \emph{Moyal-Weyl
limit}.

Another vacuum of interest is
\be
\bar X^a=X_0^a\otimes\one_n \label{XNtimes}
\ee
In this case the Moyal-Weyl limit is given by matrix valued functions on $\real^D_\theta$ and the gauge symmetry is given by unitary elements of
the algebra of $n\times n$ matrices of functions of the $X_0$. We
say that this theory has a noncommutative $U(n)$ gauge symmetry
because in the semiclassical limit it corresponds to a
nonabelian
gauge theory.

\subsection{Moyal-Weyl, gauge theory and extra dimensions
\label{se:extradim}}

Let us now consider the case in which not all dimensions
have the same significance. Split the $D$ matrices as
\be
X^a = (X^\mu,\X^i), \qquad \mu = 0,...,3,\,\, i = 1,..., n
\ee
into 4 ``spacetime'' generators $X^\mu$ which will be interpreted
as (quantized) coordinate functions, and $n$ generators $\X^i$
which are interpreted as extra dimensions. More specifically, we
consider a background (i.e.\ a solution) of the matrix model where
4 ``spacetime'' generators $X^\mu$ generate the Moyal-Weyl quantum
plane $\R^4_\theta$
\be
\bar X^\mu =X_0^\mu\otimes\one_N, \qquad \bar \X^i = 0
\ee
which satisfies
\be
\, [\bar X^\mu,\bar X^\nu] = \ii\theta^{\mu\nu}\otimes\one_N. \qquad
[\bar X^\mu,\bar \X^i] = 0 \label{barXmu}
\ee 
Here we assume $\theta^{\mu\nu} = const$ for simplicity. This 
background preserves the symmetry $SU(N)$ which commutes with 
$X^\mu$. 

Now consider small fluctuations around this solution,
\be
X^\mu = \bar X^\mu + \cA^\mu,
\qquad     \Phi^i =   \Lambda_{NC}^2\,\X^i \label{barXmufluct}
\ee
so that $X^\mu$ has dimension $length$ and $\Phi^i$ has dimension
$length^{-1}$
and can be considered (also dimensionally) a field from the four
dimensional point of view. As shown in~\cite{Steinackeroriginal}
and recalled in
Section~\eqn{se:gravitygauge}, the trace-$U(1)$ fluctuations
gives rise to the gravitational degrees of freedom which lead to
an effective (``emergent'') metric and gravity. For the sake of
the present paper we will ignore these $U(1)$ degrees of freedom
and concentrate on traceless fluctuations, assuming a flat
Moyal-Weyl background with Minkowski signature.  The remaining
$SU(N)$-valued fluctuations
\be
\cA^\mu = - \theta^{\mu\nu}A_\mu^\a(x) \otimes\la_\a
\ee
then correspond to
$SU(N)$-valued gauge fields, while the fluctuations in the internal degrees of freedom
\be
\Phi^{i} = \Phi^{i,\a}(x) \otimes\la_\a
\ee
correspond to scalar fields in the adjoint. The matrix model
action~\eqn{YM-action-1} therefore describes $SU(N)$ gauge theory on
$\R^4_\theta$ coupled to $n$ scalar fields. From now on we will drop
the $\otimes$ sign whenever there is no risk of confusion.

Noncommutative gauge theory is obtained from the matrix model
using the following basic observation
\bea
\,[\bar X^\mu + \cA^\mu,f] &=& i \theta^{\mu\nu}
(\frac{\partial}{\partial\bar x^\nu}
+ i [A_\nu,.]) f \, \equiv \, i \theta^{\mu\nu} D_\nu f .
\eea
The matrix model action~\eqn{YM-action-1} can then be written as
\bea
S_{YM} &=& \frac 1{g^2}\,\int d^4 \bar x\,
\tr\Big( G^{\mu\mu'}\,G^{\nu\nu'}\,F_{\mu\nu}\,F_{\mu'\nu'}  \nn\\
&& \quad + 2\,G^{\mu\nu}\, D_\mu\Phi^i D_\nu \Phi^i \delta_{ij}
- [\Phi^i,\Phi^j][\Phi^{i'},\Phi^{j'}] \delta_{ij} \delta_{i'j'} \nn\\
&& \quad + \bar\Psi \slashed{D} \Psi
 + \obar\Psi \Gamma_i[\Phi^i,\Psi] \Big) .
\label{YM-action-2}
\eea
This is the action of a $SU(N)$ gauge theory\footnote{The $U(1)$
components are gravitational degrees of freedom and will be
ignored here.} on $\R^4_\theta$, with effective metric given by
\be
G^{\mu\nu} = \rho\theta^{\mu\mu'}\theta^{\nu\nu'}\eta_{\mu'\nu'},
\qquad \rho = (\det\theta^{\mu\nu})^{-1/2} =: \Lambda_{NC}^4,
\label{G-def}
\ee
which satisfies $\sqrt{|G|}=1$.
Here
$F_{\mu\nu} = \partial_\mu A_\nu - \partial_\nu A_\mu
+ i[A_\mu,A_\nu]\,$ is the field strength
on $\R^4_\theta$,
\be
D_\mu \equiv \partial_\mu + i [A_\mu,.]
\ee
is the covariant derivative for fields in the adjoint,
and $\tr()$ denotes the trace over the $SU(N)$
components.
 The effective Dirac operator
is given by
\be
\slashed{D}\Psi = \Gamma_\mu \left[X^\mu, \Psi\right]
\,\sim\,  i\gamma^\mu D_\mu \Psi
\label{eq:Dirac op}
\ee
where~\cite{KlammerSteinackerfermions}
\be
\gamma^\mu = \sqrt{\rho}\,\Gamma_\nu \theta^{\nu\mu},
\qquad \{\gamma^\mu,\gamma^\nu\} = 2 G^{\mu\nu}.
\ee
The fermions have been rescaled appropriately,
and a constant shift as well as total derivatives
in the action are dropped.
Note that $g$ is now identified as the coupling constant
for the nonabelian gauge fields on $\R^4_\theta$.

\subsection{Fluctuations of the Vacuum, Emergent Gravity and Gauge Theory \label{se:gravitygauge}}

As explained above,
fluctuations of $X^a$ can be parametrized in terms of
gauge fields $A_\mu$ and scalar fields $\phi^i$ on
 $\R^4_\theta$. At first sight, this might suggest that
the action~\eqn{YM-action-2} describes $U(n)$ gauge theory on
$\R^4_\theta$. However, this interpretation is not quite correct: it
turns out~\cite{Steinackeroriginal} that the trace-$U(1)$
fluctuations of both $\cA^\mu$ and $\Phi^i$ describe gravitational
degrees of freedom which modify the geometry of $\R^4_\theta$,
defining an effective (``emergent'') geometry given by
\bea
G^{\mu\nu}(x) &=& e^{-\sigma}
\theta^{\mu\mu'}(x)\theta^{\nu\nu'}(x)g_{\mu'\nu'}(x),\nn\\
i\theta^{\mu\nu}(x) &=& i\{x^\mu,x^\nu\}\,\, \sim \,\,[X^\mu,X^\nu], \nn\\
g_{\mu\nu} &=& \eta_{\mu\nu} + \partial_\mu\phi^i \partial_\nu\phi_i,\nn\\
e^{\sigma} &=&  \sqrt{\det\theta^{\mu\nu}(x) \det g_{\mu\nu}(x)} .
\eea
The $SU(n)$-valued components of $\cA^\mu$ and $\Phi^i$ describe
nonabelian gauge fields (resp.\ scalar fields) which in the
semi-classical limit (denoted by $\sim$) couple to the effective
metric $G^{\mu\nu}(x)$. Note that the ``would-be $U(1)$ gauge
fields'' $A_\mu \one$ are absorbed in the Poisson structure
$\theta^{\mu\nu}(x) = \bar\theta^{\mu\nu}(\bar x) -
\bar\theta^{\mu\mu'}\bar\theta^{\nu\nu'} F_{\mu'\nu'}$, and
similarly the $U(1)$ degrees of freedom of $\phi^i$ are absorbed in
$g_{\mu\nu}(x)$. In the following we will ignore the gravitational
degrees of freedom and only keep the $SU(n)$-valued components of
$\cA^\mu$ and $\phi^i$, focusing on the flat Moyal-Weyl space
$\R^4_\theta$.

\section{Spontaneous symmetry breaking in extra dimensions \label{se:symmbreakextradim}}
\setcounter{equation}{0}

In this section we will present two mechanisms to break the gauge
symmetry form $SU(n)$ down to a smaller group. The first with one
constant extra dimensions, and the second with inner fuzzy
spheres. The two mechanisms are not really different because the
former can be seen as an effective model of the latter.
Only the
bosonic part will be discussed in this section as well. Both
models are somewhat analogous to a GUT-like model, where the
breaking is realized through a Higgs in the adjoint.

\subsection{Single-Coordinate Effective Breaking }
\label{sec:single-coord-break}

The mechanism how to obtain non-trivial low-energy gauge groups
and particle spectrum can be understood in a simple way as
follows. Consider a model with $\bar X^\mu$ as in
\eqn{barXmu}, but this time with a single extra dimension which
we call $\X^\Phi$ which in a suitable vacuum takes the form
\be
\langle\X^\Phi\rangle =\left(\begin{array}{ccc}
\alpha_1\one_{2}& \  & \ \\ \ & \alpha_2\one_{2} & \ \\
\ & \ & \alpha_3\one_{3} \end{array}\right) .
\label{X-phi-vac}
\ee
Here $\alpha$'s are constant\footnote{Actually it is sufficient
that at least commute with the $\bar X$'s.} quantities with the
dimensions of a length, all different among themselves.
These new coordinates are still solutions of the equations
of the motion because $[X^\mu,\langle\X^\Phi\rangle]=0$ i.e.
 $\theta^{\mu\Phi}=0$, which in turn implies
that $G^{\mu\Phi}=0$ regardless on the value of $\eta^{\Phi\Phi}$.
Therefore extra coordinate is not geometric and does not correspond
to propagating degrees of freedom from the four dimensional
point of view. The new coordinate is not invariant for the transformation
\be
\langle\X^\Phi\rangle\to U \langle\X^\Phi\rangle U^\dagger
\neq \langle\X^\Phi\rangle
\ee
for a generic $U\in SU(7)$. The traceless generators commuting
with $\langle\X^\Phi\rangle$ generate
the surviving  gauge group
$SU(2)\times SU(2) \times SU(3) \times U(1) \times U(1)$.

In the bosonic action as in Section~\ref{se:extradim} the spacetime
($\mu\nu$) part of action remains unchanged, while for the $\mu\phi$
components we obtain, in the Moyal-Weyl background,
\bea
[\bar X^\mu +\cA^\mu, \X^\Phi] &=& i\theta^{\mu\nu} D_\nu \X^\phi
=  i\theta^{\mu\nu} (\partial_\nu +  i A_\nu) \X^\phi , \nn\\
-(2\pi)^2 \Tr [X^\mu,\X^\phi] [X^{\nu},\X^\phi] \eta_{\mu\nu} &=&
\int d^4 x G^{\mu\nu} \left(\partial_\mu \X^\Phi \partial_\nu
\X^\Phi
- [A_\mu,\X^\Phi][A_\nu,\X^\Phi]\right) . \nn\\
\label{kinetic-X-A}
\eea
Note that the mixed terms $\int \partial^\mu \X^\Phi [A_\mu,\X^\Phi]
= - \frac 12 \int  \X^\phi [\partial^\mu A_\mu,\X^\Phi] = 0$
vanish, assuming the Lorentz gauge $\partial^\mu A_\mu = 0$.

Now consider the vacuum~\eqn{X-phi-vac}. Since $X^\mu$ and
$\langle\X^\Phi\rangle$ commute, this means $\langle\X^\Phi\rangle =
\mathrm{const}$ and the first term in the integral above vanish. We
can therefore separate the fluctuations of this extra dimension
which are a field, the
(high energy) Higgs field. In the action the first term is nothing
but the 
derivative of it. The second term instead is
\be
[A^\mu,\langle\X^\Phi\rangle]=\left(\begin{array}{ccc}
 0 & (\alpha_2-\alpha_1) A^\mu_{12} & (\alpha_3-\alpha_1) A^\mu_{13} \\
 (\alpha_1-\alpha_2) A^\mu_{21} & 0 & (\alpha_3-\alpha_2) A^\mu_{23} \\
 (\alpha_1-\alpha_3) A^\mu_{31} & (\alpha_2-\alpha_3) A^\mu_{32} & 0
\end{array}\right)
\ee
where we consider the block form of $A^\mu$
\be
A^\mu=\left(\begin{array}{ccc}
 A^\mu_{11} & A^\mu_{12} &  A^\mu_{13} \\
 A^\mu_{21} & A^\mu_{22} &  A^\mu_{23} \\
 A^\mu_{31} &  A^\mu_{32} & A^\mu_{33}
\end{array}\right)
\ee
Therefore~\eqn{kinetic-X-A} leads to the mass terms for the
off-diagonal gauge fields,
\be
-(2\pi)^2 \Tr [X^\mu,\langle\X^\Phi\rangle] [X^{\nu},\langle\X^\Phi\rangle] \eta_{\mu\nu}
=\int d^4 x G^{\mu\nu}
\left(\sum (\a_i-\a_j)^2 A_{\mu,ij} A_{\nu,ji} \right)
\label{A-mass-term}
\ee
which is nothing but the usual Higgs effect.
If we now assume that the differences $\alpha_i-\alpha_j$ are large, say of the grand unification scale,
it is easy to see that all non diagonal blocks of
$A^\mu$ acquire large masses
$m_{ij}^2 \sim (\a_i-\a_j)^2 $, thus effectively decoupling.

In order to approach the standard model, we will
assume the following version of the above mechanism
\be
\langle\X^\Phi\rangle=\left(\begin{array}{ccc}
\alpha_1\,\one_{2}& \  & \ \\ \ & \alpha_2\,\sigma_3 & \ \\
\ & \ & \alpha_3\,\one_{3} \end{array}\right)
\label{X-vac-3211}
\ee
with $\a_1\neq \a_2\neq \a_3$. Then the surviving traceless gauge
group is given by $SU(3) \times SU(2) \times U(1)\times U(1)\times
U(1)$, and the off-diagonal gauge fields $A_{\mu,ij}$ for $i,j$
labeling the 4 blocks acquire a mass as in~\eqn{A-mass-term}. Again,
this may simply be a crude picture of some more sophisticated
mechanism involving fuzzy spheres (branes) in extra dimensions, as
discussed below. This then comes very close to some of the proposals
how to recover the standard-model using branes and strings
stretching between them,
cf.~\cite{Antoniadis:2002qm,Blumenhagen:2006ci}.
Thus in a sense we show how such a
mechanism can be realized in the matrix model framework.

\subsection{Fuzzy Sphere Breaking \label{se:fuzzyhebreak}}

According to the splitting of the matrices into a noncommutative
spacetime $\real_\theta^4$ and ``extra'' generators $\X^i$, it is
quite natural to add
extra terms to the potential and add a potential term involving
quadratic and cubic terms
in the fluctuations $\Phi^i$ defined in~\eqn{barXmufluct}:
\bea
V_{soft}(\Phi^i) &=& 2\Tr \left(c_2 \Phi^{i}\, \Phi^{j} \delta_{ij} + i c_3
\varepsilon_{ijk} \Phi^{i}\, \Phi^{j}\, \Phi^{k} \right) \nn\\
&=& 2\Lambda_{NC}^4 \int d^4 x\
\tr\left(c_2 \Phi^{i}\, \Phi^{j} \delta_{ij} + i c_3
\varepsilon_{ijk} \Phi^{i}\, \Phi^{j}\, \Phi^{k} \right)
\label{action-S2}
\eea
From the point of view of field theory on $\R^4_\theta$,
these amount to soft (resp.\ relevant) terms  which
may (partially) break
the global $SO(n)$ symmetry, as well as supersymmetry if applicable.
In particular they may be generated upon quantization.
The full 1-loop effective potential can have more
complicated effective potentials, but for the present work
we will limit our considerations to these terms.

The bosonic part of the action~\eqn{YM-action-1} 
now becomes the following gauge theory action on $\R^4_\theta$,
\bea
S_{YM} &=& \int d^4 x\,\frac{1}{g^2} \tr  F_{\mu\nu} F_{\mu'\nu'}
G^{\mu\mu'}G^{\nu\nu'} \,\,
+ \,\, 2 \Lambda_{NC}^4\int d^4 x\,
\tr \Big( \frac{1}{g^2} G^{\mu\nu}  D_\mu \Phi^iD_\nu \Phi^i \nn\\
&& - \frac{\Lambda_{NC}^4}{2g^2}
[\Phi^i,\Phi^j][\Phi^{i'},\Phi^{j'}] \delta_{ii'}\delta_{jj'}
 + c_2 \Phi^{i}\, \Phi^{j} \delta_{ij}
+ i c_3 \varepsilon_{ijk} \Phi^{i}\, \Phi^{j}\, \Phi^{k}\Big)
\label{YM-action-2fuzzy}
\eea
with $G^{\nu\nu'}$ is as in~\eqn{G-def}. We omit here surface terms
(such as $\int d^4 x\, F_{\mu\nu}\theta_{\mu\nu}$), as well as all
trace-$U(1)$ degrees of freedom which are part of the gravitational
sector.

\paragraph{Scalar fields and spontaneous symmetry breaking.}

Assuming that the extra coordinates commute with the space time
ones, $[\X^{i},X^\mu] =[\Phi^i,X^\mu]=0$, or that in other words, in the
Moyal-Weyl vacuum they commute with the $x$'s, the above action
leads to the following equation of motion for the scalar fields
\be
\frac{2\Lambda_{NC}^4}{g^2}[\Phi^j,[\Phi^i,\Phi^{j'}]]\delta_{jj'} + 2c_2\, \Phi^i
+ \frac 32 \ii c_3 \varepsilon_{ijk}  [\Phi^{j}, \Phi^{k}] =0
\ee
This equation has as solution
\be
\langle\Phi^i\rangle= a J^i_N
\label{FS-config}
\ee
where $J^i_N$ are generators of
the $N\times N$ representation of $SU(2)$
\be
[J_N^i,J^j_N]=\ii\varepsilon_{ijk} J_N^k, \qquad
J_N^iJ_N^i=\frac{N^2-1}4 .
\ee
This solution is interpreted as fuzzy sphere
\cite{Madore:1991bw}
with radius
\be
\a = \frac a2 \sqrt{N_i^2-1}
\ee
as in \cite{AGSZ}. The equations of the motion then reduce to
\be
\frac{2\Lambda_{NC}^4}{g^2} a^2 +2c_2 -3c_3 a=0,
\ee
which generically has 2 solutions. It is important to note that
one of them
really is a global minimum of the potential for
$\Phi^i$~\eqn{YM-action-2fuzzy}:
\bea
V[\Phi^i] &=& \tr\Big(-\frac{\Lambda_{NC}^4}{2g^2} [\Phi^i,\Phi^j][\Phi^{i'},\Phi^{j'}] \delta_{ii'}\delta_{jj'}
 + c_2 \Phi^{i}\, \Phi^{j} \delta_{ij}
 + i c_3 \varepsilon_{ijk} \Phi^{i}\, \Phi^{j}\, \Phi^{k}\Big)  \nn\\
&=& - N (\frac{\a^4\Lambda_{NC}^4}{2 g^2}  + c_2 \a^2  - c_3 \a^3).
\label{fuzzypotential}
\eea
This potential has either one or two degenerate minima as a function
of $\a$, and~\eqn{FS-config} is the physical vacuum of the model.
The $SU(N)$ gauge symmetry is then broken completely by the presence
of a fuzzy sphere in the internal space.

A physically more relevant vacuum could be the one corresponding to
a ``stack'' of fuzzy spheres as proposed in \cite{AGSZ}, 
in particular\footnote{More
sophisticated versions are of course conceivable \cite{Chatzistavrakidis:2009ix}.} 
\be
\langle\Phi_i\rangle =  \begin{pmatrix}
   a_1\,J_{N_1}^i \otimes \one_2 & 0 & 0 \\
    0 & a_2\, J_{N_2}^i \otimes \sigma_3 & 0 \\
    0 & 0 & a_3\, J_{N_3}^i \otimes \one_3
\end{pmatrix} ,
\label{Z-2}
\ee
which gives an explicit realization of~\eqn{X-vac-3211}. Each of the
blocks corresponds to a fuzzy sphere $S^2_{N_i}$ with radius
$\alpha_i$. More precisely, the last block has a 3-fold
multiplicity, which can be interpreted as stack of 3 coinciding
fuzzy spheres with radius $\a_3$. If these fuzzy spheres are large,
then the fluctuations around this vacuum effectively ``see'' only
the radius of the fuzzy spheres. Thus at low energies we are very
close to the case of the previous subsection. The symmetry in the
vacuum~\eqn{Z-2} is broken down to $SU(3) \times SU(2)\times U(1)
\times U(1) \times U(1)$, which is very close to what we want. Note
that by setting e.g. $\alpha_1=\alpha_2$ and $N_1=N_2$ the symmetry
is enhanced, and more sophisticated symmetry breaking processes with
several steps (resp.\ scales) are conceivable.

An important bonus compared with standard Higgs scenarios is that
the above Higgs fields have a natural geometrical interpretation in
terms of compact fuzzy spaces. The double commutator has an
interpretation in terms of a higher-dimensional curvature, and the
additional potential~\eqn{action-S2} is cubic. This should lead to
milder renormalization properties and less fine-tuning compared with
the standard $\phi^4$ case, as observed in~\cite{AGSZ}.

\paragraph{Massive vector bosons and Higgs effect.}

The masses of the four dimensional nonabelian gauge bosons $A_\mu$
in the presence of such a fuzzy sphere vacuum were studied in
\cite{AGSZ}. The result is essentially the same as in section
\ref{sec:single-coord-break}, i.e.\ the off-diagonal components
$A_{\mu,ij}$ for $i\neq j$ acquire a mass due to the term $\tr
G^{\mu\nu} D_\mu \Phi^i D_\nu \Phi^i $, provided $(\a_i,N_i) \neq
(\a_j,N_j)$.
For the diagonal blocks,
only the $l=0$ mode of the decomposition of
$Mat(N_i) = \oplus_{l=0}^{2N_i-1} Y_{l,m}$ into
fuzzy spherical harmonics remains massless, while all higher
Kaluza-Klein modes acquire  a mass $m^2 \sim \a^2 \Lambda_{NC}^4\, l(l+1)$.
This is nothing but a
geometrical version of the usual Higgs effect.
Therefore the low-energy sector of such a fuzzy sphere
vacuum is essentially captured by the effective single-variable
description in section \ref{sec:single-coord-break}.
However, the fuzzy sphere scenario
provides a natural origin of a Higgs
potential with nontrivial minimum, which is not seen in the
single-variable description.

\section{Particle assignments, charges and symmetries}

\setcounter{equation}{0}

In this section we show how the fermions in the standard model can
be naturally accommodated in the framework of matrix models. This
is nontrivial because the  fermions in the matrix model
are necessarily in the adjoint of some basic $SU(N)$ gauge group. In a
later section we will also show how the electroweak symmetry can
be broken through a somewhat modified Higgs sector, and the Yukawa
couplings are obtained.

We start with the matrix model of Section~\ref{se:symmbreakextradim}
in $D = 4 + n$ dimensions. Thus the fermions are realized as
$D$-dimensional spinors $\Psi$ in the adjoint of $SU(N)$,
and there are
$n$ scalar fields $\Phi^i$ in the adjoint of $SU(N)$
as well as the 4-dimensional gauge
fields\footnote{Which in turn are obtained as fluctuations of the
covariant coordinates in the matrix model.} $A_\mu$.

The basic idea is to assume that the fundamental $SU(N)$ gauge group
is spontaneously broken in several steps down to the low-energy
gauge group $SU(3) \times U(1)$ of the standard model. It is natural
to assume that the various (intermediate and low-energy) gauge
groups are realized as block-diagonal subgroups of $SU(N)$. We will
show in later section how this can be realized by spontaneous
compactification on fuzzy internal spaces. In this section, we
simply assume that in a first step (at very high energy) the
block-matrix decomposition in Sect~\ref{se:symmbreakextradim} has
occurred and that therefore the symmetry is broken to $SU(3)\times
SU(2) \times U(1) \times U(1)\times U(1)$ as in~\eqn{X-vac-3211}. We
assign the fermions accordingly by the matrix
\be
\Psi= \begin{pmatrix}
  \cL_{4\times 4} & \cQ \\
  \cQ' & 0_{3\times 3}
\end{pmatrix} \label{matrixPsi-1}
\ee
Here the $4\times 4$ block $\cL$ will contain the leptons which are
color-blind, and $Q$ (resp.\ $Q'$) will contain the quarks (which we
assume to be in $(\obar 3)$ here for convenience). We drop all
fermions in the adjoint of an unbroken gauge group, i.e. we assume
that they are very massive. This is plausible if this block arises
from Kaluza-Klein modes on some fuzzy sphere as discussed above; in
principle such fermions would correspond to gauginos. We denote the
off-diagonal blocks according to~\eqn{X-vac-3211} as
\bea
\cL &=&  \begin{pmatrix}
  0_{2\times 2} & L_L \\
   L_L' & {\scriptsize \begin{array}{cc}
  0 &  e_R \\
  e_R' & 0
\end{array}} \end{pmatrix} , \nn\\
L_L &=& \begin{pmatrix}
   \tilde l_L & l_L
\end{pmatrix} , \qquad
l_L = \begin{pmatrix} \nu_L \\ e_L   \end{pmatrix}, \qquad
\tilde l_L = \begin{pmatrix} \tilde e_L \\ \tilde \nu_L
\end{pmatrix} \label{matrixPsi-2}
\eea
Here $l_L$ will be the standard (left-handed) leptons, and $e_R$ the
right-handed electron. Fields with a prime may either be related to
the unprimed ones through some conjugation, or they be independent
new fields, or they may vanish for some reason; this will be
discussed below. In particular $\tilde l$ will correspond to
additional leptons, which may be present at some energy, or which
may be null; at present the model allows them and we will keep the
term. The quarks split accordingly as
\bea
\cQ &=&  \begin{pmatrix}
  Q_L \\
  Q_R \end{pmatrix} , \nn\\
Q_L &=& \begin{pmatrix}
   u_L \\ d_L
\end{pmatrix}, \qquad
Q_R = \begin{pmatrix}
   d_R \\ u_R
\end{pmatrix}
\eea
which will again correspond to the standard quarks. This gives the
following general fermionic matrix,
\be
\Psi=\begin{pmatrix}
  0_{2\times 2} & L_L & Q_L \\
  L_L' & {\scriptsize \begin{array}{cc}
  0 &  e_R \\
  e_R' & 0
\end{array}} & Q_R \\
  Q_L' &  Q_R' & 0_{3\times 3}
\end{pmatrix}
\label{particle-assign}
\ee
The correct hypercharge, electric charge and baryon number are
then reproduced by the following traceless generators
\bea
Y &=& \begin{pmatrix}
   0_{2\times 2}  &  &  &  \\
   & - \sigma_3 & &  \\
   & &  & -\frac 13 \one_{3\times 3}
\end{pmatrix} +\frac17 \one   \label{Ygenerator}\\
Q &=& T_3 + \frac Y2 = \frac 12 \begin{pmatrix}
   \sigma_3  &  &  &  \\
   & - \sigma_3 & &  \\
   & &  & -\frac 13 \one_{3\times 3}
\end{pmatrix} +\frac1{14} \one  \label{Qgenerator}\\
B &=& \begin{pmatrix}
   0  &  &  &  \\
   & 0 & &  \\
   & &  & - \frac 13\one_{3\times 3}
\end{pmatrix} +\frac17 \one \label{Bgenerator}
\eea
which act in the adjoint.
For easy reference we display the charges $(Q,Y,B)$
of these block-matrices for the above generators:
\be
(Q,Y,B)\Psi =  \begin{pmatrix}
  0_{2\times 2} & \begin{pmatrix} (1,1,0) \\ (0,1,0) \end{pmatrix}\,
 \begin{pmatrix} (0,-1,0) \\ (-1,-1,0) \end{pmatrix} & \begin{pmatrix}
   (\frac 23,\frac 13,\frac 13) \\ (-\frac 13,\frac 13,\frac 13)
\end{pmatrix} \\[2ex]
  * & {\scriptsize \begin{array}{cc}
       0 & \qquad (-1,-2,0) \\
       *  & \qquad 0
       \end{array}}\!\!\! &  \begin{pmatrix}
   (-\frac 13,-\frac 23,\frac 13) \\ (\frac 23,\frac 43,\frac 13)
\end{pmatrix} \\[2ex]
  *  & *  & 0_{3\times 3}
\end{pmatrix}
\ee
 as it should be, omitting the obvious  lower-diagonal entries.
In particular, the charges of the exotic leptons
$\tilde l$ are those of Higgsinos.

The main result is that all particles of the standard model with the
correct quantum numbers fit naturally into this framework, based on
matrices in the adjoint of a fundamental $SU(N)$ gauge symmetry.
This is very important because the only representations which can be
realized in noncommutative gauge theory\footnote{At a fundamental
level i.e.\ without resorting to an effective Seiberg-Witten
expansion.} are fundamental, anti-fundamental and adjoint
representations of $U(N)$ gauge groups. Matrix models provides a
natural framework to study the quantization of NC gauge theory in a
non-perturbative way. In order to be free of UV/IR mixing, it
appears that only the IKKT matrix model or close relatives are
consistent at the quantum level, which contain only matrices in the
adjoint. This is a strong and predictive restriction, which
restricts the freedom in model building, with the added bonus of an
intrinsic gravity sector~\cite{Steinackeroriginal}. Perhaps the main
result of this paper is that realistic models for particle physics
appear feasible within this framework.

There is some freedom in relation to the primed fermions appearing
in the lower block of the matrix~\eqn{matrixPsi-1}, and we need to
understand the relation among the upper and lower diagonal blocks.
It is likely that a supersymmetric version of this model can be
built, and in that framework some of zeros of the matrix can be
filled by supersymmetric partners, and the additional leptons
$\tilde l$ could be identified with Higgsinos.

Let us discuss some possibilities how fermions
may arise in the off-diagonal blocks.
We first need to understand the relation between
the upper-diagonal and the lower-diagonal components. We note that the
upper and lower triangles in the matrix are exchanged under hermitean
conjugation, which is part of charge conjugation.
Therefore, as we will see in detail in the following,
the role of primed and unprimed elements are exchanged in
$\bar\Psi$. There are three obvious choices for the primed fermions.

\begin{enumerate}
\item
If $\Psi = \Psi^C$ is a Majorana-Weyl fermion in the fundamental
matrix model, then the lower-diagonal components are related
with the upper-diagonal ones directly by charge conjugation. This choice may be natural in the presence of 10 dimension, in which case it is possible to have fermions which are both Majorana and Weyl.

\item One can set the primed fermions equal to zero, so that
    $\Psi$ is an upper triangular matrix. The lower part of the
    matrix will appear in $\bar\psi$, which will be  lower
    triangular. This choice has several advantage, as will see
    in the following, but it seems ``ad hoc'', without an
    explanation at the present. This may be related to
    the presence of a magnetic
    flux~\cite{Steinacker:2007ay}.

\item If both upper and lower-diagonal components are
    non-vanishing and not related via conjugation, the model is
    non-chiral, corresponding to a mirror model.  Then has to
    explain why each single sector has an independent
    cancellation of anomalies, which would be canceled by the
    mirrors anyway; this of course apart form consideration on
    the presence of dark matter which are still distant from the
    present state of the art.

\end{enumerate}

In the latter two case the lower-triangular case can be seen as an
instance of the presence of fermion doubling, which is known
phenomenon in noncommutative geometry~\cite{LMMS,AC2M2}. At this
stage it is really a matter of taste if one prefers to eliminate the
lower triangle of the matrix setting it to zero, or to keep it as a
mirror world. With the former choice one is setting to zero a sector
which is in principle present, but which can give unwanted
couplings.

The correct chirality assignment is put in by hand here.
There is also a slot
which could naturally accommodate additional leptons $\tilde l_L$.
Notice also that the scheme is naturally suited for supersymmetry,
since all particles and fields
arise from matrices in the adjoint of $SU(N)$.

The full $U(N)$ model is certainly free of anomalies.
After symmetry breaking,
the $U(1)_B$ gauge symmetry may turn out to be
anomalous, as it often does in string theory~\cite{Kiritsis},
and we assume that the corresponding gauge boson becomes massive
through some version of the Green-Schwarz mechanism.
Furthermore the additional leptons $\tilde l$ lead to
an anomaly unless their lower-diagonal partners
$\tilde l$ are also present; this strongly suggest that
$\tilde l$ should be set to zero.

These ambiguities indicate that an additional mechanism is required
to single out the correct physical result. In particular, it is very
interesting that the above scheme is very similar to constructions
in string theory~\cite{Antoniadis:2002qm,Blumenhagen:2006ci,Kiritsis},
where the standard model is realized in
terms of 4 stacks of branes with exactly the above gauge groups, and
particles realized as strings stretched between these branes. The
latter correspond precisely to the off-diagonal blocks, and there
seems to be a correspondence between the possibilities indicated
above and the different versions of this construction in string
theory. This suggests that additional structures such as
intersecting branes should be considered in the matric model
framework. This is probably possible, and e.g. branes with fluxes
were recently realized in~\cite{Steinacker:2007ay}.
In particular, fuzzy orbifolds \cite{orbifolds} appear to
realize the above structures in a chiral model. We will not
investigate these in the present paper.

Here we do not claim to have the final answer, rather we want to
point out possible directions which should be pursued elsewhere. We
take this similarity with string theory as additional encouragement.
However, we want to stress that our approach offers advantages over
string theory, simply because the matrix model is a very specific
and predictive framework. For example, the branes realized as fuzzy
spheres are naturally obtained as stable minima of the
potential~\eqn{fuzzypotential}. Furthermore, this result shows
clearly that there is no obstacle to describe (an extension of) the
standard model within the framework of noncommutative gauge theory.
The mechanism is applicable to models which are expected to be
well-defined at the quantum level, in particular the IKKT
model~\cite{IKKT}.

\section{Electroweak breaking}

Now we show how electroweak symmetry breaking might be realized in
this framework. To explain the idea we will first present a
simplified version where the Higgs is realized in terms of a single
extra coordinate (resp.\ scalar) field. In section
\ref{se:EWbreakfuzzy} we then discuss a more elaborate version
involving extra coordinates (resp.\ Higgs fields), which form an
``electroweak'' fuzzy sphere. This is again not intended as a
realistic model, but it shows that suitable Higgs potential can
naturally arise within the present framework.

\subsection{Electroweak Higgs and Yukawa coupling.}

Higgs field connects the left with the right sectors of leptons, and
is otherwise colour blind, it is therefore natural to consider,
along the lines of Sect.~\ref{sec:single-coord-break}, another extra
coordinate which will have to necessarily be off-diagonal. The
following matrix has the correct characteristics:
\be
\X^\phi=\Lambda_{NC}^{-2}\left(\begin{matrix}
  0_{2\times 2} & \phi & 0_{2\times 3} \\
  \phi^\dagger & 0_{2\times 2} & 0_{2\times 2} \\
  0_{3\times 2} &  0_{3\times 2} & 0_{3\times 3}
\end{matrix}\right)
\label{Higgs-coord}
\ee
where we use again the notations of
section~\ref{sec:single-coord-break} and consider the extra variable
$\X$, its vacuum expectation value and the fluctuations which are a
physical field. The Higgs $\phi$ is a $2 \times 2$ matrix which is
actually composed of two doublets (which form the Higgs content of
the minimal supersymmetric standard model), i.e. two scalar doublets
with opposite hypercharges:
\be
\phi = \begin{pmatrix} \tilde \varphi, & \varphi
\end{pmatrix} \label{two-higgs}
\ee
The vacuum expectation value of $\phi$ is an off-diagonal matrix:
\be
\langle\phi\rangle=\left(\begin{matrix} 0 & v\\ \tilde v & 0\end{matrix}\right)
\ee
All other components
(possibly even some of the components of $\phi$) are assumed
to be very massive, e.g. due
to the commutator with the high-energy breaking discussed before.

Now consider the fermionic part of the action~\eqn{YM-action-1},
which can be written on $\R^4_\theta$ in the form ~\eqn{YM-action-2}.
The part involving $X^\mu$ gives the usual Dirac action as
in~\eqn{YM-action-2},
and the part involving $\X^\phi$
yields the Yukawa couplings
\be
S_Y = \Tr \obar\Psi \gamma_\phi [\X^\phi,\Psi]
\ee
giving mass to the fermions.
Here $\gamma_\phi$ is an
extra-dimensional gamma matrix corresponding to $\X^\phi$.
We write
\be
\obar \Psi = \Psi^\dagger \gamma_0=
\begin{pmatrix}
  0_{2\times 2} & \obar{L_L'} & \obar{Q_L'} \\
  \obar L_L & {\scriptsize \begin{array}{cc}
  0 &  \obar e_R' \\
  \obar e_R & 0
\end{array}} & \obar{Q_R'} \\
  \obar Q_L &  \obar Q_R & 0_{3\times 3}
\end{pmatrix}
\ee
Then the full Yukawa term without any omissions or further assumptions is
\bea
S_Y &=& \Tr \Big(-\obar{L'} \gamma_\phi \left({\scriptsize
\begin{array}{cc}
  0 &  e_R \\
  e_R' & 0
\end{array}}\right) \phi^\dagger
                - \obar{Q_L'} \gamma_\phi Q_R' \phi^\dagger \nn\\
 && + \obar L_L \gamma_\phi \phi \left({\scriptsize \begin{array}{cc}
  0 &  e_R \\
  e_R' & 0
\end{array}}\right)
+ \left({\scriptsize \begin{array}{cc}
  0 &  \obar e_R' \\
 \obar e_R & 0
\end{array}}\right) \gamma_\phi  (\phi^\dagger L - L' \phi)
- \obar{Q_R'}\gamma_\phi Q_L' \phi \nn\\
 && + \obar Q_L \gamma_\phi \phi Q_R
+ \obar Q_R \gamma_\phi \phi^\dagger Q_L \Big)
\label{Yukawa-general}
\eea
We now impose $\gamma_\phi=\gamma_5$, which is natural since in this
way the five dimensional Clifford algebra is closed, and
\be
\gamma_5 L_L = + L_L, \qquad \gamma_5 Q_L = + Q_L, \qquad \gamma_5
Q_R = - Q_R \qquad \gamma_5 e_R=-e_R
\ee
with this assumption and  we see that the couplings come to be the
correct ones. Only opposite chiralities couple in such a Yukawa
term.

The construction is quite solid and works in all three case for the
primed and unprimed fermions. in the case for which the primed
fermions vanish only a few terms will survive. In the case of
Majorana fermions the couplings are the ones needed to give Dirac
masses to Majorana fermions. In the case of mirror fermions there is
no coupling among mirror and ordinary fermions, so that the mirror
world effectively decouples.

The extra fermion doublet $\tilde l$ does not couple with the
remaining leptons with the option of setting the primed fermions to
zero. If the primed sector is the conjugate sector of Majorana
fermions there is a coupling of $e_R$ and $\tilde e_L$ which may
cause problems. Note that in this model the masses and the Yukawa
couplings of leptons and quarks come to be the same and there is no
way to differentiate them. A breaking with fuzzy spheres discussed
below give more structure to the extra dimensions
and may create differences.

Assuming
\be
\langle\phi\rangle=\begin{pmatrix} 0 & v\\ \tilde v & 0
\end{pmatrix} \label{vacuumphi}
\ee
with $v$ and $\tilde v$ real, we get that the quark contribution to
the action is
\bea
&&\bar{\mathcal{Q}} \gamma_5\begin{pmatrix} 0 & \phi\\ \phi^\dagger &
0\end{pmatrix} \mathcal{Q} - \bar{\mathcal{Q}}' \gamma_5\mathcal{Q}'
\begin{pmatrix} 0 & \phi^\dagger \\ \phi &
0\end{pmatrix}=\nonumber\\&&-v(\bar d_R  d_L - \bar d_L  d_R -\bar
d_R'  d_L'  +\bar d_L'  d_R') - \tilde v( \bar u_R  u_L - \bar u_L
u_R
 -\bar u_R'  u_L' + \bar u_L'  u_R')\nonumber\\
\eea
the lepton part of the action is instead
\be
\bar{\mathcal L}\gamma_\phi [\mathcal L,\phi]=-\tilde v(-\bar e_L
e_R  + \bar e_R e_L + \bar e_L' e_R  - \bar e_R e_L') - v(-\bar
e_R\tilde e_L' + \bar e _R'\tilde e_L -\bar{\tilde e}_L e_R'
+\bar{\tilde e}_L' e_R)
\ee
Note that with  choice~\eqn{vacuumphi} the leptons of the $\tilde l$
doublet do not have mass terms, but have spurious coupling to the
ordinary leptons. It is possible to set them to zero in this scheme,
but then we get that the mass of the electron is the same as the one
of the up quark. Note also that if we relax the reality requirement
on the $v$'s then the coefficients of $\bar e_L e_R$ and $\bar e_R
e_L$ are complex conjugate of each other, and the same will hold for
quarks. There is no problem in setting the primed fermions to zero,
the hermitean conjugated appear naturally because of $\bar \Psi$,
and if we set $\tilde l=0$ then there also is no problem for
Majorana spinors. Mirror fermions have again no problems, except
that it is still not clear the mechanism to give then large mass. It
is still too early for a complete analysis of the various choices
for the couplings since we are not yet at the stage to be building a
completely realistic model. For example there are no generations,
nor different couplings for the different gauge groups, and this
points to the necessity of the refinement of the model.

\subsection{Electroweak Breaking by fuzzy sphere \label{se:EWbreakfuzzy}}

Consider the fuzzy sphere breaking at high energies
(at the GUT scale, say) described in
section~\ref{se:fuzzyhebreak} and in particular the stack of fuzzy
sphere breaking described in~\eqn{Z-2}. The residual gauge symmetry
in this case is $SU(3)\times SU(2) \times U(1)_Q \times U(1)_Y \times
U(1)_B$.
We will later discuss the
splitting into  $SU(4) \times SU(3) \times U(1)$
corresponding to the $4\times 4$ lepton block the
$3\times 3$ plus color block.

As in section~\ref{se:fuzzyhebreak}, we assume that there are
additional quadratic and cubic terms as in~\eqn{action-S2} in the
effective potential
\be
V_{H}(\X_i^\phi) = \Tr \left(-\frac 1{g^2}\, [\X_i^\phi,\X_j^\phi]^2
 + c_2 \X_{i}^\phi\, \X_{j}
  ^\phi\delta^{ij}
+ i c_3 \varepsilon^{ijk} \X_{i}^\phi\, \X_{j}^\phi\, \X_{k}^\phi \right)
\label{action-EW}
\ee
at the electroweak scale. A possible minimum of this
potential is given by the following
fuzzy sphere\footnote{Here we indicate only the relevant $4 \times 4$
  block in square brackets and drop the color blocks.} $S^2_{EW-I}$:
\bea
\langle\X^\phi_1\rangle &=&  \a_H \left[\begin{matrix}
   \sigma_3 & 0  \\
    0 & \sigma_3
\end{matrix}\right] = \one\otimes \sigma_3, \nn\\
\langle\X^\phi_2\rangle &=&  \a_H  \left[\begin{matrix}
    0  & \sigma_2  \\
    \sigma_2 & 0
\end{matrix}\right] = \sigma_1\otimes \sigma_2, \nn\\
\langle\X^\phi_3\rangle &=&  \a_H  \left[\begin{matrix}
    0  & \sigma_1  \\
    \sigma_1 & 0
\end{matrix}\right] = \sigma_1\otimes \sigma_1 ,
\label{S2-I}
\eea
This breaks the symmetry $SU(3)\times SU(2) \times U(1)_Q \times
U(1)_Y \times U(1)_B$ of~\eqn{Z-2} down to $SU(3) \times U(1)_Q
\times U(1)_B$, and we have indeed achieved the desired electroweak
symmetry breaking.
Observe that$ \langle\X^\phi_2\rangle$ and $\langle\X^\phi_3
\rangle$ are very similar to the two Higgs doublets $H, \tilde H$ in
the MSSM, with an additional 3rd Higgs $\langle\X^\phi_1\rangle$ in
the diagonal blocks. Since the off-diagonal blocks are assumed to
have definite chirality as in the standard model, this diagonal
Higgs does not contribute to the Yukawa couplings. However, it does
contribute to the mass of the $W^\pm$ and $Z$ bosons. This will be
discussed below.

Alternatively, if we start from a vacuum
\be
\langle\Phi_i\rangle =  \begin{pmatrix}
   a_1\,J_{N_1}^i \otimes \one_4 & 0 \\
     0 & a_3\, J_{N_3}^i \otimes \one_3
\end{pmatrix} ,
\label{Z-4}
\ee with $SU(4) \times SU(3)
\times U(1)$ symmetry, then the single fuzzy sphere~\eqn{S2-I} is
not sufficient, since it commutes with the generators
${\X'}^\phi_i=Q\X^\phi_i$,
\bea
\langle{\X'}^\phi_1\rangle &=&  \a_H' \left[ \begin{matrix}
   \one & 0  \\
    0 & -\one
\end{matrix}\right] =  \sigma_3\otimes\one, \nn\\
\langle{\X'}^\phi_2\rangle &=&  \a_H' \left[ \begin{matrix}
    0  & -\ii\sigma_1  \\
    \ii\sigma_1 & 0
\end{matrix}\right] = \sigma_2\otimes \sigma_1, \nn\\
\langle{\X'}^\phi_3\rangle &=&  \a_H' \left[ \begin{matrix}
    0  & \ii\sigma_2  \\
    -\ii\sigma_2 & 0 &
\end{matrix} \right] =-\sigma_2\otimes \sigma_2 ;
\label{S2-Iprime}
\eea
The above
six matrices close a $SO(4)$ Lie Algebra 
\bea
{}[\langle\X^\phi_i\rangle,\langle\X^\phi_j\rangle] &=& - 2\ii \varepsilon_{ijk} \a_H\langle\X^\phi_k\rangle \nonumber\\
{}[\langle{\X}^\phi_i\rangle,\langle{\X'}^\phi_j\rangle] &=& 2\ii\varepsilon_{ijk}\a'_H \langle{\X'}^\phi_k\rangle \nonumber\\
{}[\langle{\X'}^\phi_i\rangle,\langle{\X'}^\phi_j\rangle] &=&
-2\ii\varepsilon_{ijk} \frac{{\a'_H}^2}{\a_H}
\langle{\X}^\phi_k\rangle \nonumber\\
\frac1{\a_H^2}\langle\X^\phi_i\rangle \langle\X^\phi_i\rangle
&=&\frac1{{\a'_H}^2}\langle{\X'}^\phi_i\rangle
\langle{\X'}^\phi_i\rangle =\one_{4\times 4} \label{FS-one}
\eea
The two commuting $SO(3)$ algebras are then
\be
\langle{\X^\pm}^\phi_i\rangle=\frac1{2\a_H}\langle\X^\phi_i\rangle\pm\frac1{2\a_H'}\langle{\X'}^\phi_i\rangle
\ee
and they represent two fuzzy spheres which commute with $Q$ (and of
course the identity).

Thus in that case we can achieve the desired symmetry breaking down
to $SU(3) \times U(1)_Q \times U(1)_B$ using these two fuzzy
spheres. There may be important differences between these scenarios
depending on the energy scales of these spheres, and more detailed
work is required before claiming any direct phenomenological
relevance. In any case, our main point is that it seems
feasible to obtain a (near-) realistic extension of the standard
model by these or similar mechanisms. The essential ingredients,
notably the stacks of various fuzzy spheres,  are
similar to string-theoretical constructions of (extensions of the)
standard-model using branes in extra dimensions. Similar 
ideas are also used in \cite{AGSZ,orbifolds}.

The coordinates of the fuzzy spheres couple with the fermions in the
action~\eqn{YM-action-2} via the term $\obar\Psi
\Gamma_i[\Phi^i,\Psi]$, with the $\Phi$'s proportional to the $\X$'s
as in~\eqn{barXmufluct} and the $\gamma$'s of the internal
dimensions represented as diagonal matrices in the $7 \times 7$
gauge matrix space. The Yukawa couplings for the fuzzy
sphere~\eqn{S2-I} are then (omitting the proportional factor
$\alpha_H$):
\bea
\tr\obar\Psi[{\X_1}^\phi,\Psi]&=& -\obar d_L d_L+ \obar d_R d_R +
 2 \obar e_R e_R + \obar u_L u_L  - \obar u_R u_R+ 2 \obar\nu \nu - 2 \obar{\tilde\nu}
 \tilde\nu\nonumber\\&&
+\obar d_L' d_L'- \obar d_R' d_R' -
 2 \obar e_R' e_R' - \obar u_L' u_L'  + \obar u_R' u_R'- 2 \obar\nu' \nu' + 2 \obar{\tilde\nu}'
 \tilde\nu'\nonumber\\
\tr\obar\Psi[{\X_2}^\phi,\Psi]&=&-\ii \obar d_R d_L + \ii \obar d_L
d_R + \ii \obar e_L e_R + \ii \obar u_R u_L  - \ii \obar u_L  u_R
\nonumber\\&&
 - \ii \obar{\tilde e_L}' e_R - \ii \obar
d_R' d_L' + \ii\obar d_L' d_R' +\ii \obar e_R' \tilde e_L - \ii
\obar e_R'  e_L  -\ii  \obar e_R e_L + \ii \obar e_R \tilde e_L'
\nonumber\\&&- \ii \obar{\tilde e}_L e_R' + \ii \obar e_L' e_R' +
\ii \obar u_R' e_L'  - \ii \obar u_L' u_R'
\nonumber\\
\tr\obar\Psi[{\X_3}^\phi,\Psi]&=&\obar d_R d_L +  \obar d_L d_R +
\obar e_L e_R + \obar u_R u_L  + \obar u_L  u_R \nonumber\\&&
 -  \obar{\tilde e_L}' e_R -  \obar
d_R' d_L' -\obar d_L' d_R' +\obar e_R' \tilde e_L -  \obar e_R' e_L
+  \obar e_R e_L - \obar e_R \tilde e_L' \nonumber\\&&- \obar{\tilde
e}_L e_R' - \obar e_L' e_R' - \obar u_R' e_L' - \ii \obar u_L'
u_R'
\eea
while if one considers the pair of spheres~\eqn{FS-one} (setting
$\alpha_H=\alpha'_H=1$ for simplicity) one obtains:
\bea
\tr\obar\Psi[{\X^+_1}^\phi,\Psi]&=& \obar e_L e_L + \obar{\tilde
e}_L \tilde e_L + \obar e_R e_R +\obar u_L u_L - \obar u_R u_R +2
\obar\nu \nu \nonumber\\&& -\obar e_L' e_L' - \obar{\tilde e}_L'
\tilde e_L' - \obar e_R' e_R' -\obar u_L' u_L' + \obar u_R' u_R'
-2
\obar\nu' \nu'\nonumber\\
\tr\obar\Psi[{\X^+_2}^\phi,\Psi]&=&\ii \obar u_R u_L - \ii \obar
u_L u_R - \ii \obar {\tilde e}_L'e_R  + \ii \obar e_R {\tilde e}_L
+ \ii \obar e_R {\tilde e}_L' - \ii \obar {\tilde e}_L e_R' + \ii
\obar u_R' u_L' - \ii \obar u_L' u_R'
\nonumber\\
\tr\obar\Psi[{\X^+_3}^\phi,\Psi]&=&\obar u_R \obar u_L + \obar u_L
\obar u_R - \obar {\tilde e}_R' e_R + \obar e_R' {\tilde e}_L -
\obar e_R {\tilde e}_L' + \obar {\tilde e}_L e_R' - \obar u_R'
u_L' - \obar u_L' u_R'
\nonumber\\
\tr\obar\Psi[{\X^-_1}^\phi,\Psi]&=&-\obar d_L d_L   + \obar d_R
d_R - \obar e_L e_L - \obar {\tilde e}_L {\tilde e}_L + \obar e_R
e_R - 2 \obar{\tilde \nu} {\tilde \nu} +\nonumber\\&& \obar d_L'
d_L' - \obar d_R' d_R' + \obar e_L' e_L'  + \obar {\tilde e}_L'
{\tilde e}_R' - \obar e_R' e_R' + 2 \obar {\tilde \nu}'
{\tilde\nu}' \nonumber\\
\tr\obar\Psi[{\X^-_2}^\phi,\Psi]&=&-\ii \bar d_R d_L  + \ii \bar
d_L d_R - \ii \bar e_R e_L + \ii \bar e_L e_R - \ii \bar d_R' d_L'
+ \ii \bar d_L' d_R' - \ii \bar e_R' e_L' + \ii \bar e_L' e_R'
 \nonumber\\
\tr\obar\Psi[{\X^-_3}^\phi,\Psi]&=& \bar d_R d_L+ \bar d_L d_R +
\bar e_R e_L + \bar e_L e_R - \bar d_R' d_L' - \bar d_L' d_R'
-\bar e_R' e_L' - \bar e_L' e_R'
\eea
The couplings which appear are all ``reasonable'', meaning that
they are either Majorana or Dirac masses, or coupling among the
primed particles or the spurious leptons. Setting all of these to
zero we obtain:
\bea
\tr\obar\Psi[{\X^+_1}^\phi,\Psi]&=& \obar e_L e_L + \obar e_R e_R
+\obar u_L u_L - \obar u_R u_R +2 \obar\nu \nu \nonumber\\
\tr\obar\Psi[{\X^+_2}^\phi,\Psi]&=&\ii \obar u_R u_L - \ii \obar
u_L u_R
\nonumber\\
\tr\obar\Psi[{\X^+_3}^\phi,\Psi]&=& \obar u_R \obar u_L + \obar
u_L \obar u_R
\nonumber\\
\tr\obar\Psi[{\X^-_1}^\phi,\Psi]&=&-\obar d_L d_L  + \obar d_R
d_R - \obar e_L e_L + \obar e_R e_R \nonumber\\
\tr\obar\Psi[{\X^-_2}^\phi,\Psi]&=&-\ii \bar d_R d_L  + \ii \bar
d_L d_R - \ii \bar e_R e_L + \ii \bar e_L e_R
 \nonumber\\
\tr\obar\Psi[{\X^-_3}^\phi,\Psi]&=& \bar d_R d_L+ \bar d_L d_R +
\bar e_R e_L + \bar e_L e_R
\eea
These are the couplings of the standard model in the absence of
right handed neutrinos. Some of these terms may vanish depending
on the specific chirality assignment, as discussed before.
As it is the model does not allow for
different masses, apart from
some freedom afforded by the tuning of $\alpha_H$ and $\alpha_H'$.

%
%
%
%
Note that the remaining eight generators of $Mat(4,\complex)$ do
not commute with these two fuzzy spheres, thus the gauge symmetry
is indeed broken to $Q$ and the generators of colour and baryon
number. According to what we explained in the previous sections
this implies that they are massive.

It is worthwhile to elaborate in some detail the explicit form of
the low-energy electroweak Higgs. Consider first the
vacuum without fluctuations. Using
\bea
\langle\X^\phi_i\rangle \langle\X^\phi_j\rangle \delta^{ij} &=& \a_H^2 , \nn\\
\varepsilon^{ijk} \langle\X_{i}^\phi\rangle\,
\langle\X_{j}^\phi\rangle\, \langle\X_{k}^\phi\rangle &=& -2\ii
\a_H^3
\eea
so that the effective potential for $\a_H$ becomes
\be
V_{H}(\langle\X_i^\phi\rangle) = \Tr \left(\frac 4{g^2}\, \a_H^4
+ c_2 \a_H^2 + 2 c_3 \a_H^3 \right) .
\label{action-EW-value}
\ee
This has a non-trivial minimum in $\a_H \neq 0$
provided $c_3 \neq 0$ or $c_2<0$,
and the sign of $\a_H$ depends on the sign of $c_3$.
Note that these terms have a geometrical interpretation
in terms of
field strength on $S^2_N$, leading to some protection from
quantum corrections \cite{AGSZ}.

The VEV's of $\X_{2}^\phi$ and $\X_{3}^\phi$ contains the expected
degrees of freedom of the two Higgs doublets $\phi$ as
in~\eqn{two-higgs}, parametrizing one complex scalar. This is as in
the MSSM, however the two doublet are related to each other. They
become independent in the presence of the second fuzzy
sphere~\eqn{S2-Iprime}. The VEV of $\X_{1}^\phi$ contains scalar
degrees of freedom which are in the adjoint of the electroweak
$SU(2)$, with the same VEV. This is different from the standard
model and should have observable signatures. We therefore obtain an
interesting geometrical interpretation of these scalar Higgs fields.

Now consider fluctuations around this vacuum. Again, these
fluctuations contain fluctuations of the two Higgs doublets $\phi$
as in~\eqn{two-higgs}, and also fluctuations of $\X_{1}^\phi$ which
is in the adjoint of the electroweak $SU(2)$. More generally,
fluctuations on the fuzzy sphere can be interpreted as scalar (resp.
gauge fields) on $S^2_N$.

\paragraph{Vector boson masses.}

As explained in section \ref{se:fuzzyhebreak},
the vector bosons corresponding to the
$4\times 4$ leptonic block will acquire particular mass terms
in the presence of these electroweak fuzzy spheres.
For example, $\X^\phi_1 $ will contribute a mass
proportional to $\a_H^2$ to the $W$ bosons.

Furthermore, note that $\X'^\phi_1 $ breaks $SU(4)$ into $SU(2)
\times SU(2)$, which seems quite appealing; this suggests that
$\a_H'$ should have higher scale than $\a_H$, on the other hand
then $\X'^\phi_{2,3}$ lead to EW symmetry breaking which is
strange. This suggests some interplay between the two spheres.

\section{Conclusions and Outlook}

In this paper we have shown how the matrix model which gives rise to
noncommutative spaces and emergent gravity can also accommodate a
gauge theory with the features of the standard model. We have seen
that a simple solution with extra (non-propagating) dimensions
contains all the necessary fundamental fermionic degrees of freedom,
with a few extra particles which can be set to zero without
prejudice to the model. Also the basic gauge symmetries can be
accommodated and, with the use of extra dimensions, the pattern of
symmetry breakings can be substantially reproduced. The breaking
happen in two stages, first some sort of grand-unification breaking,
and then the electroweak breaking. Both stages can be accomplished
either with the presence of a simple (effective) extra matrix
dimension, or with the use of fuzzy spheres. The former mechanism
can be considered an effective version of the latter.

There are several gaps in the constructions, and several lines of
developments which hopefully can fill the gaps. The list of
shortcomings includes the fact that there are some extra $U(1)$
symmetries, the lack of generations, and
the fact that couplings do not
differentiate between fermions and bosons. Clearly the
solutions discussed here are not
phenomenologically viable, but we find rather inspiring the fact
that we have a semi-realistic matrix model from which gravity and
gauge theories of a realistic kind emerge naturally. Among the lines
of development there is the possibility to have a supersymmetric
version of the model. This type of matrix model is in fact
well suited for supersymmetric extensions,
the most notable example being the IKKT model \cite{IKKT}.
Moreover, the introduction of additional geometrical structures such
as intersecting branes and orbifolds
\cite{orbifolds} is likely to resolve at least some of these problems,
in particular the issue of chirality.
Another line of development is a better understanding of the
connections among the extra dimensions in the guise of fuzzy spheres
and he results obtained in string and brane theory.
This supports the hope that the framework of matrix models
might be suitable to approach the goal of a
consistent quantum theory of fundamental interactions
including gravity.

\subsection*{Acknowledgements} H.S. would like to thank 
Athanasios Chatzistavrakidis and George Zoupanos
for many discussions and related collaboration.
FL~would like to thank the
Department of Estructura i
Constituents de la materia, and the Institut de Ci\`encies del
Cosmos, Universitat de Barcelona for hospitality. His work has been
supported in part by CUR Generalitat de Catalunya under project
2009SGR502.
The work of H. S. was supported in part by the FWF project P18657
and in part by the FWF project P21610.

\end{document}